\documentclass[iop]{emulateapj}
\usepackage{natbib}
\usepackage{ulem}
\usepackage[usenames]{color}
\slugcomment{Accepted for publication in the ApJ: June 9, 2011}

\newcommand       \msun        	{$M_{\odot}$}
\newcommand       \zsun      	{$Z_{\odot}$} 
\newcommand       \lsun      	{$L_{\odot}$}

\newcommand	     \myr              {$M_{\odot}$~yr$^{-1}$}
\newcommand       \mic        	 {$\mu$m}

\newcommand	\az				{AzTEC--3}

\newcommand	\lbol			{$L_{bol}$}
\newcommand	\mdust			{$M_{dust}$}
\newcommand	\mgas			{$M_{gas}$}
\newcommand	\mdyn			{$M_{dyn}$}
\newcommand	\mstar			{$m_{\star}$}
\newcommand	\tmax			{$t_{age}(max)$}
\newcommand	\massav			{$\left< m \right>$}

\begin{document}

\title{STAR AND DUST FORMATION ACTIVITIES  IN AzTEC-3,\\ A STARBURST GALAXY AT z = 5.3}

\author{Eli Dwek\altaffilmark{1}, Johannes G. Staguhn\altaffilmark{1,2}, Richard G. Arendt\altaffilmark{1,4}, Peter L. Capak\altaffilmark{3}, Attila Kovacs\altaffilmark{6}, 
Dominic Benford\altaffilmark{1}, 
 Dale Fixsen\altaffilmark{1,4}, Alexander Karim\altaffilmark{5}, Samuel Leclercq\altaffilmark{7}, Stephen F. Maher\altaffilmark{1,8}, Samuel H. Moseley\altaffilmark{1}, Eva Schinnerer\altaffilmark{5}, and Elmer H. Sharp\altaffilmark{1,9}}

\altaffiltext{1}{Observational Cosmology Lab, Code 665, NASA Goddard Space Flight Center, Greenbelt, MD 20771, USA; {\it eli.dwek@nasa.gov}}
\altaffiltext{2}{The Henry A. Rowland Department of Physics and Astronomy, Johns Hopkins University, 3400 N. Charles Street, Baltimore, MD 21218, USA}
\altaffiltext{3}{Spitzer Science Centre, 314-6 California Institute of Technology, 1200 E. California
Blvd., Pasadena, CA 91125, USA}
\altaffiltext{4}{CRESST, University of Maryland -- College Park, College Park, MD 20742, USA}
\altaffiltext{5}{Max-Planck-Institute f\"ur Astronomie, K\"onigstuhl 17, D-69117, Heidelberg, Germany}
\altaffiltext{6}{University of Minnesota, 116 Church St SE, Minneapolis, MN 55414, USA}
\altaffiltext{7}{Institut de RadioAstronomie Millim\'etrique, 300 rue de la Piscine, 38406 Saint
Martin d'H\`eres, France}
\altaffiltext{8}{Science Systems and Applications, Inc., 10210 Greenbelt Rd, Suite 600, Lanham, MD 20706, USA}
\altaffiltext{9}{Global Science \& Technology, Inc., 7855 Walker Drive, Suite 200, Greenbelt, MD 20770, USA}

\begin{abstract}
Analyses of high-redshift ultraluminous infrared (IR) galaxies traditionally use the observed optical to submillimeter spectral energy distribution (SED) and estimates of the dynamical mass as observational constraints to derive the star formation rate (SFR), the stellar mass, and age of these objects. An important observational constraint neglected in the analysis is the mass of dust giving rise to the IR emission. In this paper we add this constraint to the analysis of \az. Adopting an upper limit to the mass of stars and a bolometric luminosity for this object, we construct different stellar and chemical evolutionary scenarios, constrained to produce the inferred dust mass and observed luminosity before the associated stellar mass exceeds the observational limit. We use the P\'EGASE population synthesis code and a chemical evolution model to follow the evolution of the galaxy's SED and its stellar and dust masses as a function of galactic age for seven different stellar initial mass functions (IMFs). We find that the model with a Top Heavy IMF provided the most plausible scenario consistent with the observational constraints. In this scenario the dust formed over a period of $\sim 200$~Myr, with a SFR of $\sim 500$~\myr. These values for the age and SFR in \az\ are significantly higher and lower, respectively, from those derived without the dust mass constraint. However, this scenario is not unique, and others cannot be completely ruled out because of the prevailing uncertainties in the age of the galaxy, its bolometric luminosity, and its stellar and dust masses.
A robust result of our models is that all scenarios require most of the radiating dust mass to have been accreted in molecular clouds. 
Our new procedure highlights the importance of a multiwavelength approach, and of the use of dust evolution models in constraining the age and the star formation activity and history in galaxies.
\end{abstract}

\section{INTRODUCTION}
The origin of the dust in ultraluminous high redshift galaxies with redshifts $z \gtrsim 4$ is still unresolved, with supernovae (SNe), evolved AGB stars, the winds around active galactic nuclei (AGN), or growth in molecular clouds as possible sources \citep{elvis02,morgan03,maiolino04a,dwek07b,valiante09,dwek11,michalowski10a,gall11a,gall11b}. Further complicating the origin of dust in some objects is the energy source powering the IR emission. An IR luminosity powered by luminous massive stars will suggest a high present SFR, favoring a SN origin for the dust. An IR luminosity that is powered by an active galactic nucleus (AGN) suggests a low present SFR and may favor the delayed injection of dust by AGB stars that may have formed during an earlier phase of much higher star formation activity \citep{valiante09,dwek11}.
 
No such ambiguity exists for \az, an ultraluminous IR galaxy located at redshift $z = 5.3$, that was first detected at submillimeter wavelengths by \cite{scott08}. It is a radio-dim source \citep{younger07} with no X-ray counterpart \citep{capak11}, and part of a young cluster, representing an overdensity of early massive galaxies at the same redshift \citep{capak11}.  Optical to submillimeter observations summarized in \cite{capak11} show that the energy output is dominated by its far-infrared (IR) emission with luminosities between $\sim (0.6-2)\times 10^{13}$~\lsun. 
The lack of any radio or X-ray counterparts suggests that this luminosity is powered by a 1800~\myr\ starburst \citep{riechers10,capak11}. The total CO-inferred gas mass is $\sim 5\times10^{10}$~\msun, which will be depleted in less than $\sim 30$~Myr at this SFR.
Such energy output suggests the presence of a large amount of dust, which had to be created (at least) in less than $\sim 1.1$~Gyr, the age of the universe at that redshift. \az\ therefore falls in the category of extreme starburst galaxies with large dust masses ($\gtrsim 10^8$~\msun) described by \cite{michalowski10b} and \cite{robson04}.  

In this paper we present a new comprehensive approach for studying the star formation properties and the origin of dust in \az. The approach is quite general and can be applied to any local or high-redshift galaxy. We first use the UV to far-IR SED of the galaxy to determine its bolometric luminosity and dust mass. Its SFR depends on the adopted IMF, and we generate a series of models relating the bolometric luminosity to the stellar mass and galactic age for different stellar IMFs (Section~2). In Section~3 we use the observed UV-optical (UVO) spectrum, the UVNIR SED, and dynamical limits on the stellar mass to constrain the age of the galaxy from observations. We then generate different scenarios for the evolution of the dust, characterized by different IMFs, and examine which scenario is capable of producing the inferred dust mass in \az\ within the allowed age limit, that is, before the concurrent stellar mass production exceeds the stellar mass limit (Section~4). Our results are discussed and summarized in Section~5.

In all our calculations we adopt a flat $\Lambda$CDM cosmology, with a baryonic density parameter $\Omega_b=0.044$, a total matter (dark+baryonic) density parameter of $\Omega_m=0.27$, a vacuum energy density $\Omega_{\Lambda}=0.73$, and a Hubble constant of $H_0 = 70~$~km~s$^{-1}$~Mpc$^{-1}$ \citep{spergel07}. 

\section{PROPERTIES OF \az}
  \subsection{Bolometric Luminosity and Dust Mass}

\begin{figure}[b]
  \centering
  \includegraphics[width=3.5in]{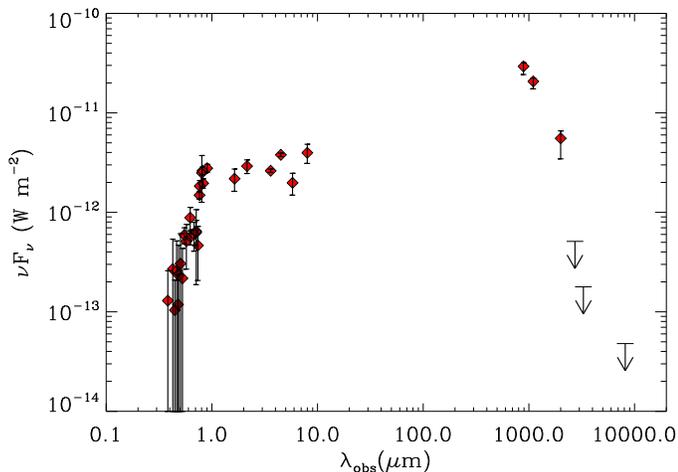} 
   \caption{{\footnotesize The observed visual to submm fluxes from \az, calculated for a redshift of $z=5.3$, are plotted as a function of wavelength. }}
\label{fig1}
\end{figure}

\begin{figure}[htbp]
  \centering
  \includegraphics[width=3.0in]{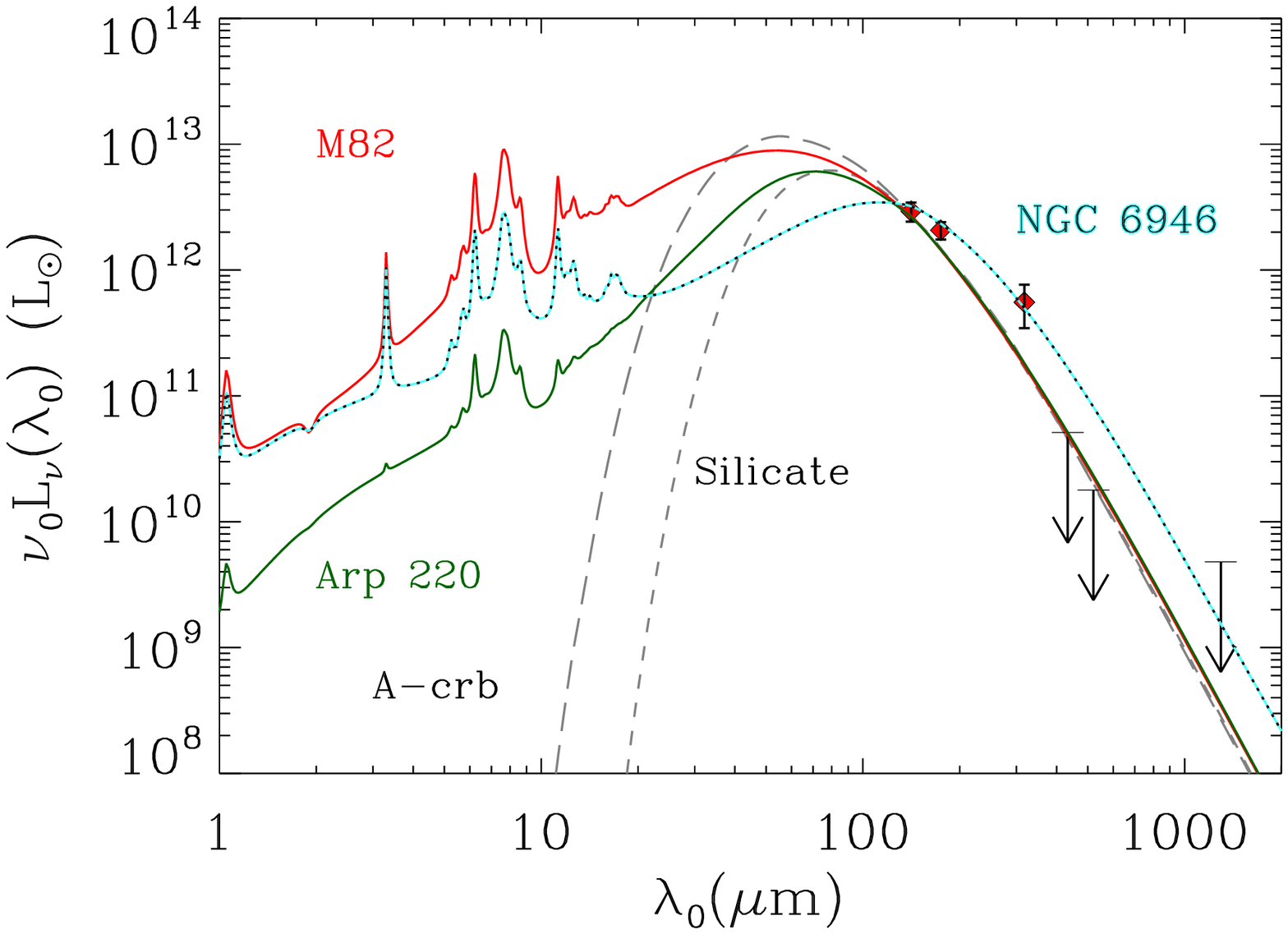} 
  \hspace{0.1in}
    \includegraphics[width=3.0in]{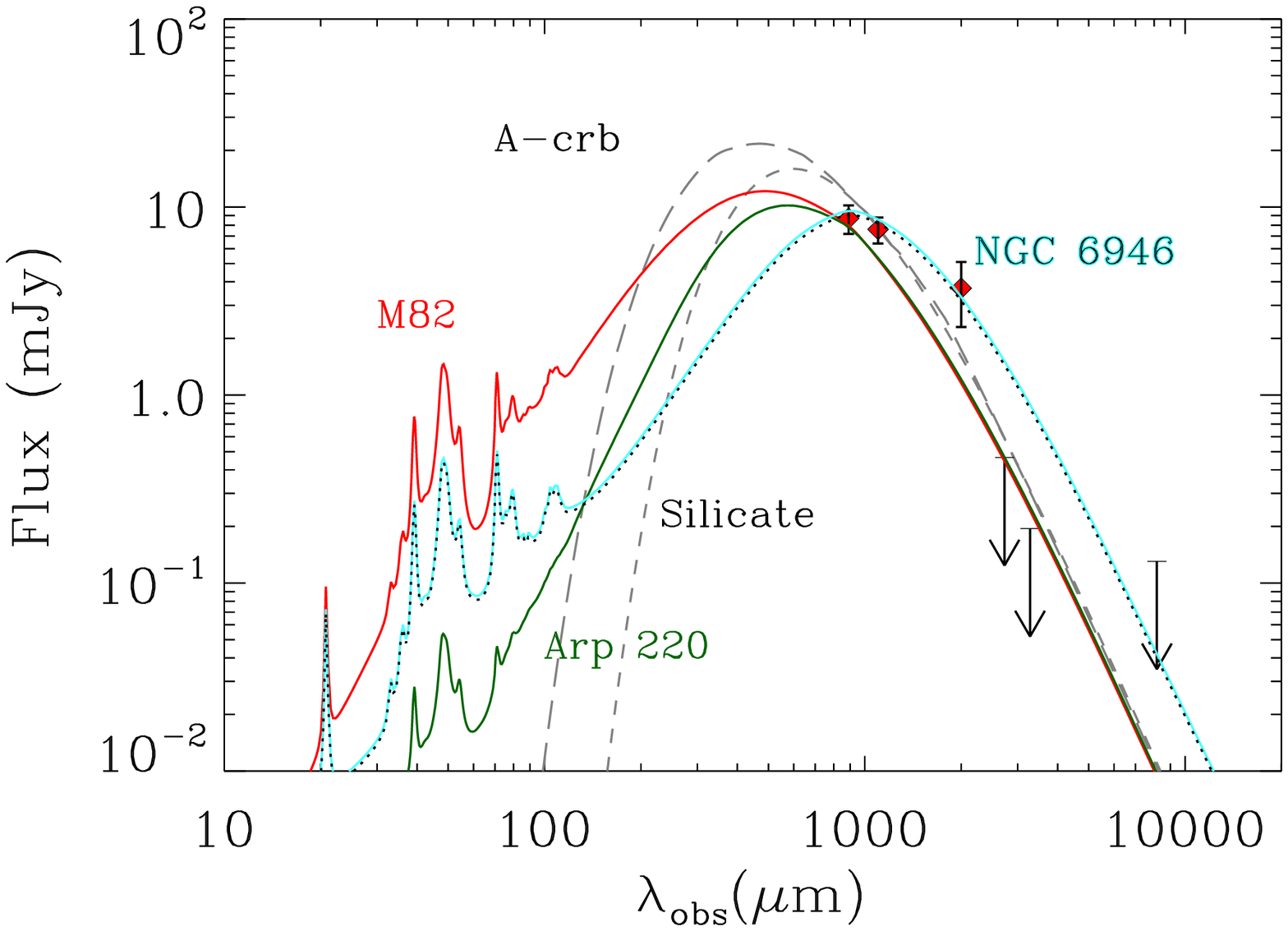}
  \vspace{0.1in}
  \caption{{\footnotesize The intrinsic far-IR luminosity (top panel) and the observed specific intensity (bottom panel) of \az\ are plotted as a function of the rest frame and observed wavelengths, respectively. Also plotted are template spectra of local galaxies and single-temperature fits of dust grains of different compositions to the observed spectrum. Data and references are given  in Table \ref{submm_obs}. }}
\label{irspec}
\end{figure}

Figure \ref{fig1} depicts the visible to far-IR flux, $\nu\, F_{\nu}$, from the galaxy calculated for a redshift of $z=5.3$. The observed fluxes were taken from \cite{capak11}.
The figure shows that the energy output from the galaxy is dominated by the stellar radiation that is absorbed and reradiated by dust at IR wavelengths. Observed IR flux densities and their 1$\sigma$ uncertainties are listed in Table~\ref{submm_obs}. Figure \ref{irspec} shows the far-IR galaxy's SED in its rest frame (left panel) and the observer's frame (right panel).  Also shown in the Figure are fits of the spectra of individual dust species to the observations. In order to determine the uncertainties in the derived dust temperatures, dust masses, and IR luminosities, we constructed 1000 realizations of the observed spectrum. The flux at each wavelength was randomly chosen from a normal distribution centered around the nominal flux.  The derived quantities for the different dust compositions are listed in Table~\ref{dust_tml}. The Fe and graphite dust spectra are similar to that of the silicate dust, and were left out of the figure for sake of clarity.
Dust masses vary greatly from $0.3\times10^9$~\msun, if all the dust in \az\ is in the form of amorphous carbon, to $3.2\times10^9$~\msun, if it consists of only silicate dust. Because of the skin effect, the absorptivity of metallic Fe depends on the the grain radius, and the mass presented in the table was calculated for a radius of 0.33~\mic\ for which the mass absorption coefficient, $\kappa$, was maximum at the wavelength of 174.6~\mic. This choice of $\kappa$ minimizes the mass of Fe dust required to fit the observed fluxes.  Since \az\ is not likely to contain only pure Fe, silicate, or  carbon-type dust, any dust consisting of a mixture of these elements will vary between between these mass estimates.  

\begin{deluxetable}{ccccc}
\tablewidth{0pt}
\tablewidth{0pt}
\tablecaption{Summary of Submillimeter Observations of \az \tablenotemark{1}}
\tablehead{
\colhead{$\lambda_{obs}$~(\mic)} &
\colhead{$F_{\nu}(\lambda)$ (mJy)} &
\colhead{$\sigma$ (mJy)} & 
\colhead{$\lambda_0$~(\mic)\tablenotemark{1}} & 
\colhead{$L_{\nu}(\lambda_0) (L_{\odot}$~Hz$^{-1})$}  
}
\startdata
     890&        8.7&      				1.5&      141.3&       1.158 \\
      1100&      7.6&      				1.2&      174.6&       1.011 \\
      2000&      3.7 &                    1.4 &     317.5&       0.492 \\
      2730&     0.47\tablenotemark{2}&     \nodata &      433.3&     0.0412     \\
      3280&     0.20\tablenotemark{2}&     \nodata&      520.6&     0.0173 \\
      8165&    0.087\tablenotemark{2}&    \nodata&      1296.0&     0.0116 \\
\enddata
\tablenotetext{1}{Restframe wavelengths and specific luminosities were calculated for a redshift of $z=5.3$.}
\tablenotetext{2}{3-$\sigma$ upper limits.}
\label{submm_obs}
\end{deluxetable}

\begin{deluxetable*}{lcccc}
\tablewidth{0pt}
\tablewidth{0pt}
\tablecaption{Derived Temperatures, Masses, and IR Luminosities for Different Dust Compositions and Galaxy Templates}
\tablehead{
\colhead{Composition} &
\colhead{$T_d$ (K)} &
\colhead{$M_d~(10^9M_{\odot})$} & 
\colhead{$L_{IR}~(10^{13}L_{\odot})$} & 
\colhead{$\kappa$ (cm$^2$ g$^{-1})$\tablenotemark{1}}  
}
\startdata
Fe\tablenotemark{2} 	& 32.4$\pm$ 6.1	& $1.4^{+1.2}_{-0.7}$ & $0.6^{+0.4}_{-0.2}$ & 16.5 \\
Graphite 			& 27.5$\pm$4.5	& $2.2^{+4.6}_{-2.2}$ & $0.5^{+0.3}_{-0.2}$ & 17.6 \\
Silicate 			& 28.5$\pm$ 4.3	& $3.2^{+2.6}_{-1.4}$ & $0.6^{+0.4}_{-0.2}$ & 10.5 \\
A-carbon 			& 46.9$\pm$14.8	& $0.3^{+0.35}_{-0.17}$ & $1.1^{+1.7}_{-0.67}$ & 28.0 \\
\hline\
M 82 template		& \nodata	& $2.6\pm 0.9$ & $1.7\pm 0.6$ & \nodata \\
Arp 220 template 	& \nodata	& $1.4\pm 0.8$ & $0.8\pm 0.5$ & \nodata \\
NGC 6946 template 	& \nodata	& $18\pm 8$ & $0.6\pm0.3$ & \nodata \\
\enddata
\tablenotetext{1}{The mass absorption coefficient at wavelength $\lambda = 174.6$~\mic.}
\tablenotetext{2}{Fe mass was calculated for a grain radius of 0.33~\mic. See text for details.}
\label{dust_tml}
\end{deluxetable*}
\vspace{0.1in}

To get a more realistic presentation of the mixture of dust compositions and size distribution in \az\  we also fitted  its spectrum with that of two local starburst galaxies: M82, and Arp220, and a normal star forming galaxy NGC6946. The spectra of these galaxies were obtained by using population synthesis models to derive their stellar emission, chemical evolution models to derive their metallicity and dust composition, and radiative transfer models including the stochastic heating of dust to derive their IR spectrum and dust mass \citep{galliano08a}.  Transferred to a redshift of $z=5.3$, the IR luminosities and dust masses for all three galaxies were derived from their normalization factors, obtained from the least-squares fit of their spectra to the \az\ observations. The spectra depicted in the Figure represent the total IR emission from dust residing in HII and photodissociation regions. Both starburst galaxies, M82 and Arp220, provided a very good fit to the data. Dust masses varied from 1.4 to $2.6\times10^9$~\msun, and IR luminosities ($\sim 3-1000$~\mic) from 0.79 to $1.7\times 10^{13}$~\lsun. In contrast, the dust in NGC6946 is too cold, and its far-IR emission exceeds the 3-$\sigma$ upper limits on the underlying continuum of the detected CO($J=5\rightarrow 4$) and CO($J=6\rightarrow 5$) lines \citep{riechers10}. This is evident in the inferred dust mass for this galaxy, which is in excess of $\sim 10^{10}$~\msun. 

For sake of being definitive we adopt throughout this paper a dust mass, \mdust, of $2.0\pm1.0\times10^{9}$~\msun, representing the average value between M82 and Arp220. Our choice of average dust mass derived from these template galaxies is more realistic than any choice based on the assumption that all the IR emission is produced by a single dust specie. The average IR luminosity of these two galaxies is $1.0\times10^{13}$~\lsun\ and since the bolometric luminosity of \az\ is dominated by the IR emission component we adopt a bolometric luminosity, \lbol, that is equal to $1.0\times10^{13}$~\lsun.

~
\\ ~

  \subsection{The Star Formation Rate}
\begin{figure}[t]
  \centering
  \includegraphics[width=3.5in]{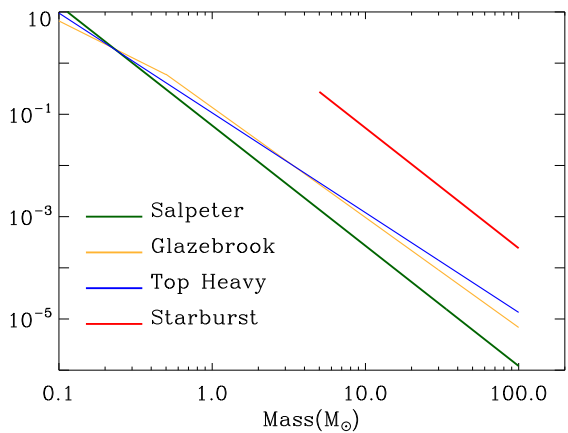}
  \vspace{0.1in}
  \caption{{\footnotesize The four stellar IMF discussed in detail in the paper as a function of stellar mass. The functional forms of these and the other IMFs are presented in Table~\ref{imf}.}}
\label{imf_plot}
\end{figure}

The far-IR luminosity provides a good measure of the star formation activities in galaxies \citep{kennicutt98a}. In a galaxy with ongoing star formation the bolometric luminosity and stellar mass increase with time because of the cumulative contribution from long-lived low-mass stars. Derived star formation rates depend therefore on the assumed star formation history of the galaxy. Furthermore, the relation between the luminosity and star formation rate depends on the adopted stellar IMF. Massive stars radiate much more efficiently than low mass ones ($L \propto M^{3.3}$, in solar units) so that an IMF weighted more heavily towards massive stars will require a lower SFR for a given observed luminosity. 
To explore the dependence of the SFR on the stellar IMF, we used seven different functional forms to characterize the latter. The parameters characterizing the IMFs are given in Table \ref{imf}, and Figure~\ref{imf_plot} depicts the functional form of 4 select IMFs. Table~\ref{imf} also includes two characteristic values of the different IMFs: \massav, which is the IMF-averaged mass, and \mstar, which is the mass of stars that need to be formed to produce one Type~II SN event. With these definitions, the stellar birthrate (yr$^{-1}$) is given by SFR/\massav, and the SN rate by SFR/\mstar.

\begin{deluxetable*}{lccccccccc}
\tablewidth{0pt}
\tablewidth{0pt}
\tablecaption{Characteristics of the Stellar IMFs\tablenotemark{1}}
\tablehead{
\colhead{IMF} &
\colhead{$\alpha_1$} &
\colhead{$\alpha_2$} & 
\colhead{$\alpha_3$} & 
\colhead{M$_1$} & 
\colhead{M$_2$} & 
\colhead{M$_3$} &
\colhead{M$_4$} &
\colhead{$\left<m\right>$\tablenotemark{2}} &
\colhead{m$_{\star}$\tablenotemark{3}}
}
\startdata
Glazebrook & 1.5 & 2.15 & \nodata & 0.1 & 0.5 & 100 & \nodata & 0.72 & 70.0\\
Kroupa     & 1.3 & 2.3 & \nodata & 0.1 & 0.5 & 100 & \nodata & 0.64   & 89.6\\
Mass Heavy & 2.35 & \nodata & \nodata & 1.0 & 100 & \nodata & \nodata & 3.1 & 52.7\\
Paunchy    & 1.0 & 1.7 & 2.6 & 0.1 & 0.5 & 4 & 100 &  1.1    & 66.4\\
Salpeter   & 2.35 & \nodata & \nodata & 0.1 & 100 & \nodata & \nodata & 0.35 & 134.7\\
Starburst  & 2.35 & \nodata & \nodata & 5.0 & 100 & \nodata & \nodata & 12.8 & 24.4\\
Top Heavy  & 1.95 & \nodata & \nodata & 0.1 & 100 & \nodata & \nodata & 0.78 & 55.4
\enddata
\tablenotetext{1}{All masses are in \msun.  Stellar IMFs are given by:  \\
 $\qquad \phi(m)  \propto 	 M^{-\alpha_1}$ for $M_1 < M < M_2$; \\
 $\qquad \phi(m)  \propto 	 M^{-\alpha_2}$ for $M_2 < M < M_3$;  and\\
 $\qquad \phi(m)  \propto 	 M^{-\alpha_3}$ for $M_3 < M < M_4$ }
 \tablenotetext{2}{The IMF averaged stellar mass.} 
  \tablenotetext{3}{The mass of all stars born per SN event \citep{dwek11}} 
\label{imf}
\end{deluxetable*}

 \begin{figure}[t]
  \centering
      \includegraphics[width=3.5in]{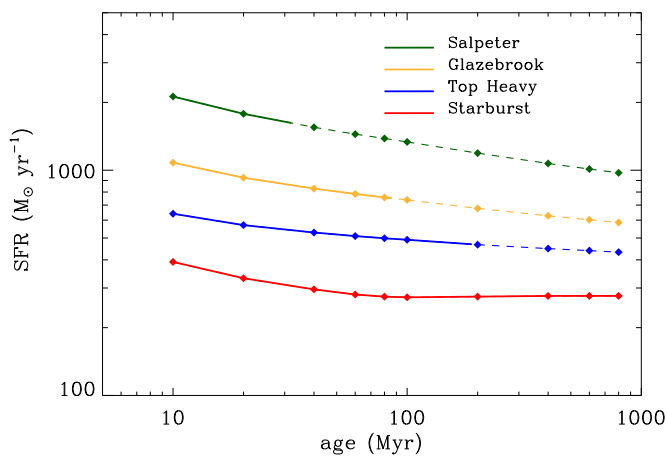}
  \caption{{\footnotesize The SFR required to produce a bolometric luminosity of $L_{bol}=1\times10^{13}$~\lsun\ at a given galactic age is depicted as a function of age for four different IMFs. For each IMF, the bold colored line depicts the allowed range of galaxy ages before the model violates the constraint on the stellar mass (see Section~3.3).  }}
\label{sfr}
\end{figure}  

To determine the SFR and the mass of stellar objects as a function of galactic age we ran the P\'EGASE population synthesis code developed by \cite{fioc97} for the seven different stellar IMFs, listed in Table~\ref{imf}, for a constant SFR of 1~\myr. 
Table~\ref{lumSB_sfr} list the corresponding bolometric luminosities and stellar masses  at different galactic ages. For example, for the adopted $L_{bol} = 1\times10^{13}$~\lsun, the table shows that for  a Glazebrook IMF the SFR is $\sim 735$~\myr\ and the total mass of stars is $\sim 5.6\times10^{10}$~\msun\ when the galaxy reaches an age of 100~Myr. At the same age, a starburst IMF will require a SFR of only $\sim 270$~\myr\ to produce the same luminosity, and will have created  $\sim 7.5\times10^{9}$~\msun\ of stars. Figure~\ref{sfr} depicts these results in graphic form,  showing the constant SFR required to produce a bolometric luminosity of $1\times10^{13}$~\lsun\ at a given age for 4 different IMFs. The bold section in each line depicts the allowed range of galaxy ages before the model violates the constraint on the stellar mass, a topic discussed in detail in Section~3.3 below. We emphasize that this figure does {\it not} represent the evolution of the SFR with time. For each stellar IMF, it depicts the value of the constant SFR needed to produce a bolometric luminosity of $1\times10^{13}$~\lsun\ at the assumed galactic age. The required SFR is lower for older systems because a larger fraction of \lbol\ arises from the accumulation of lower mass stars over the galaxy's lifetime.

\begin{deluxetable}{lcccc}
\tablewidth{0pt}
\tablewidth{0pt}
\tablecaption{Starburst Luminosities and Stellar Masses\tablenotemark{1}}
\tablehead{
\colhead{Starburst age (yr)} &
\colhead{10$^7$} & 
\colhead{10$^8$} & 
\colhead{10$^9$} & 
\colhead{10$^{10}$}
}
\startdata
 Glazebrook &  & & & \\
\hspace{0.2in} $L_{bol} \ (L_{\odot})$ &   9.26e+09 & 1.36e+10 & 1.74e+10 & 2.20e+10 \\
\hspace{0.2in} $M_{star} \ (M_\odot)$ &  1.03e+07 & 7.61e+07 & 5.89e+08 & 4.50e+09 \\
\hline
Kroupa & & & &  \\
\hspace{0.2in} $L_{bol} \ (L_{\odot})$ &  7.33e+09 &  1.13e+10  & 1.54e+10 &  2.08e+10 \\
\hspace{0.2in} $M_{star} \ (M_\odot)$ &  1.05e+07  & 8.06e+07  & 6.40e+08 &  4.92e+09  \\
\hline
Mass Heavy & & &  & \\
\hspace{0.2in} $L_{bol} \ (L_{\odot})$&    1.20e+10 &  1.90e+10 &  2.65e+10 &  3.25e+10   \\
\hspace{0.2in} $M_{star} \ (M_\odot)$&    1.02e+07 &  6.70e+07 &  3.80e+08 &  1.15e+09 \\
\hline
Paunchy & & & &  \\
\hspace{0.2in} $L_{bol} \ (L_{\odot})$&    7.83e+09 &  1.43e+10 &  2.05e+10 &  2.53e+10 \\
\hspace{0.2in} $M_{star} \ (M_\odot)$&   1.05e+07 &  7.64e+07 &  4.92e+08 &  3.08e+09  \\
\hline
Salpeter & & &  & \\
\hspace{0.2in} $L_{bol} \ (L_{\odot})$&    4.71e+09 &  7.49e+09 &  1.06e+10 &  1.49e+10 \\
\hspace{0.2in} $M_{star} \ (M_\odot)$&    1.07e+07 &  8.76e+07 &  7.56e+08  & 6.46e+09 \\
\hline
Starburst & & &  & \\                 
\hspace{0.2in} $L_{bol} \ (L_{\odot})$&    2.55e+10  & 3.67e+10 &  3.62e+10 &  3.41e+10 \\
\hspace{0.2in} $M_{star} \ (M_\odot)$&    9.20e+06  & 2.76e+07 &  3.17e+07  & 7.18e+07 \\
\hline
Top Heavy & & &  & \\
\hspace{0.2in} $L_{bol} \ (L_{\odot})$&    1.56e+10  & 2.04e+10 &  2.34e+10 &  2.56e+10 \\
\hspace{0.2in} $M_{star} \ (M_\odot)$&   9.72e+06  & 6.23e+07  & 4.45e+08  & 3.32e+09 
\enddata
\tablenotetext{1}{Entries are calculated for a constant SFR of $1~M_{\odot}~yr^{-1}$.}
\label{lumSB_sfr}
\end{deluxetable}

\section{THE AGE OF \az}

 \subsection{Spectral Constraints on the Age}
The preceding section illustrated the importance of knowing the galactic age for determining the star formation, and the stellar and elemental enrichment histories of a galaxy. The UVO spectrum can contain important information for determining this age. Figure~1 in \cite{capak11} presents the full resolution spectrum of \az, obtained with the Keck~II telescope. The spectrum shows Si~II and C~IV stellar absorption features which are indicative of the presence of a population of young O and B stars. In a single burst of star formation the spectral contribution from young stars should diminish with time. The presence of these features therefore led them to the conclusion that \az\ should be less than $\sim 30$~Myr old.
However, in an ongoing star formation scenario, the stellar population will always contain young O and B stars, and their spectral features will only fade at much later times as they are gradually overwhelmed by the featureless continuum from the less massive stars. 

 \begin{figure}[htbp]
  \centering
  \includegraphics[width=3.0in]{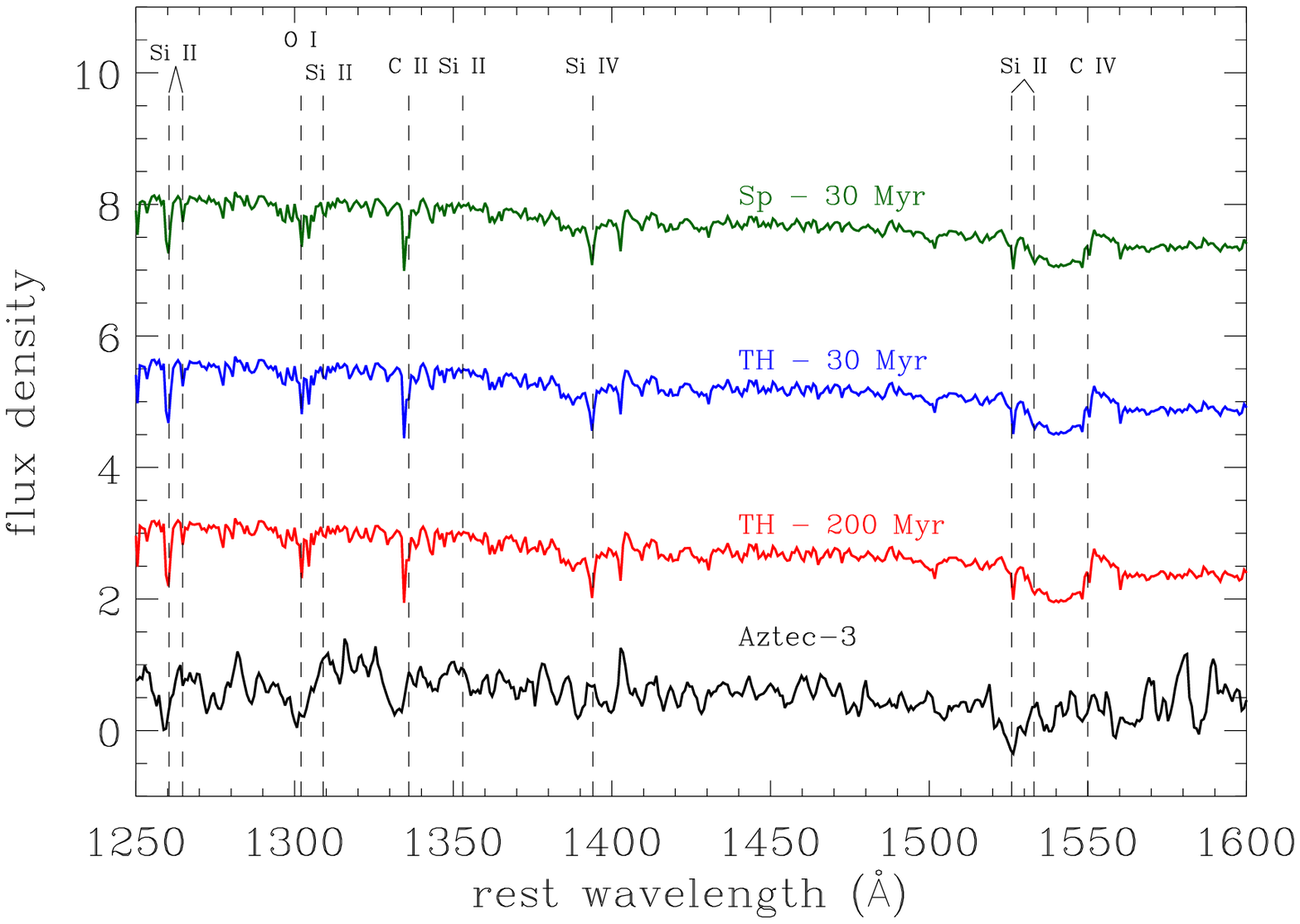} 
  \hspace{0.1in}
    \includegraphics[width=3.0in]{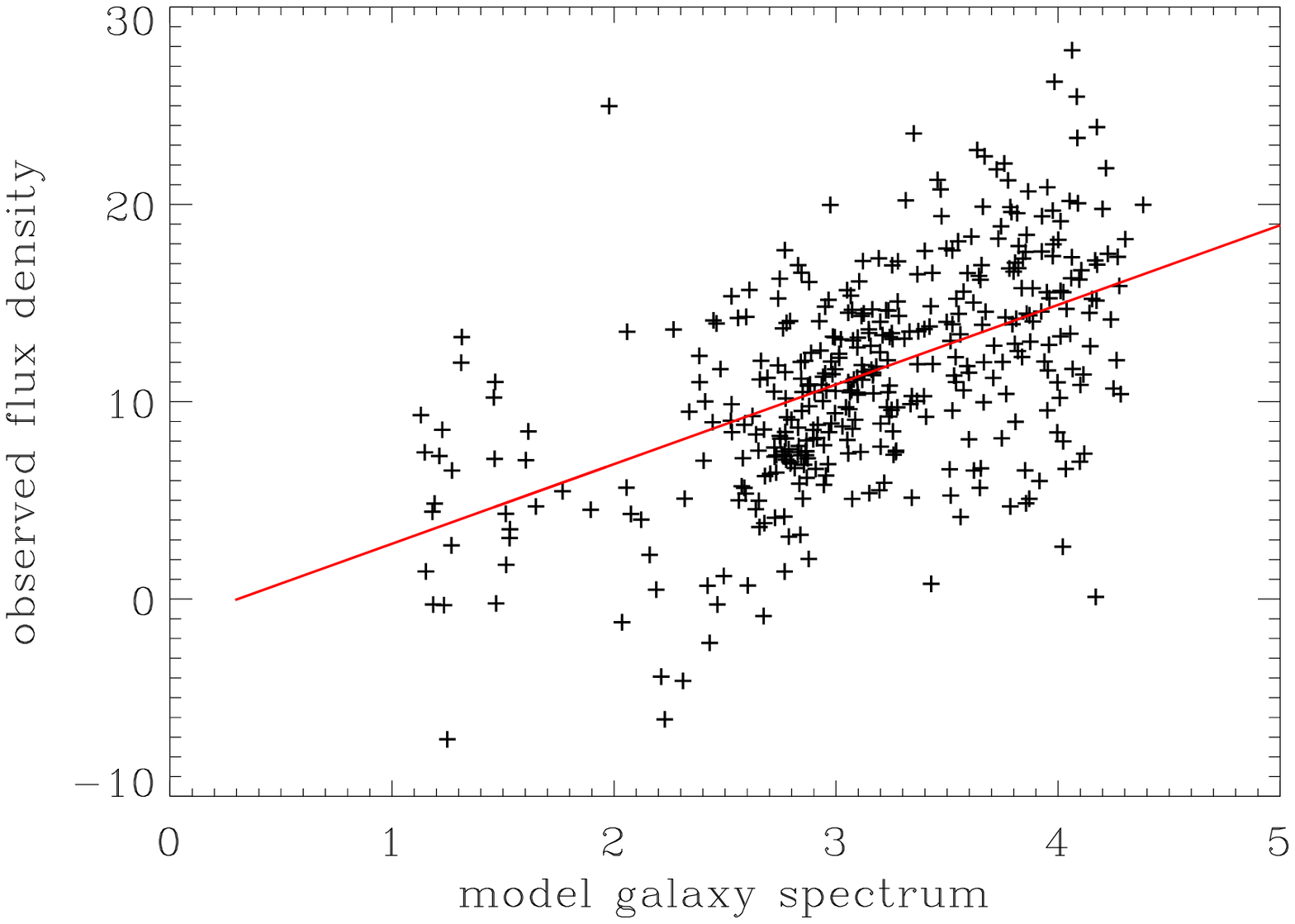}
  \vspace{0.1in}
  \caption{{\footnotesize
           {\bf Top panel}: The observed Keck~II spectrum of \az\ \citep{capak11} is compared to high-resolution UV spectra generated by Starburst99. The observed spectrum has been redshifted by $z=5.3$ and rebinned to the 0.75~\AA\ resolution of the model spectra. Model spectra were calculated assuming ongoing constant star formation rates, and are depicted for two stellar IMFs: Salpeter (Sp) and Top-Heavy (TH). To facilitate the comparison between the models and observations the spectra were normalized and shifted with respect to each other. The figure also depicts the location of important stellar lines. The observed \az\ spectrum is more complex since it also contains interstellar absorption lines. The main point of this panel is to illustrate that the observed spectrum cannot be used as an effective determinator of the age of \az.  {\bf Bottom panel}: The correlation between the observed and the model spectrum. The colored line depicts the best fit line to the correlation. The synthetic spectra are essentially identical for all models depicted in the figure (except for a scaling factor), yielding an identical degree of correlation between the models and observations, regardless of the age of the system. }}
\label{keck}
\end{figure}  

To examine if the observed spectrum can be used to determine the age of the galaxy we constructed synthetic UV spectra for constant star formation scenarios and different stellar IMFs (TH=Top Heavy; Sp=Salpeter) using Starburst99 \citep{leitherer99,vazquez05,leitherer10}. Figure~\ref{keck} compares the synthetic spectra with the observed one.  The figure shows that the TH model spectra have not evolved significantly in shape between the ages of 30 and 200~Myr. Their total intensity has, of course evolved, an effect not shown in the figure because of the offset. They are also identical to the Sp model at 30~Myr. The simulations show that when star formation is an ongoing process, the spectrum does not evolve significantly because even a very small number of O/B stars will dominate the UV emission. The \az\ spectrum shows some of the same stellar absorption features present in the synthetic spectra, but contains in addition features caused by interstellar absorption, a process that is not included in the Starburst99 models. The right panel of the figure shows the correlation between the observed spectrum and that of one of the models. Except for a constant offset caused by the evolution of the total intensity of the spectrum, the correlation plot is identical for all models.

The similarity of the \az\ spectrum to the model spectra, and the fact that the model spectra are essentially identical even though they represent different epochs and were generated with different IMFs, shows that the observed $\sim 1200 - 1600$~\AA\ spectrum cannot be used to constrain the age of the galaxy when star formation is an ongoing process.

 \subsection{SED Contraints on the Age: The Age Degeneracy}
\begin{figure*}[p]
  \centering
  \includegraphics[width=5.0in]{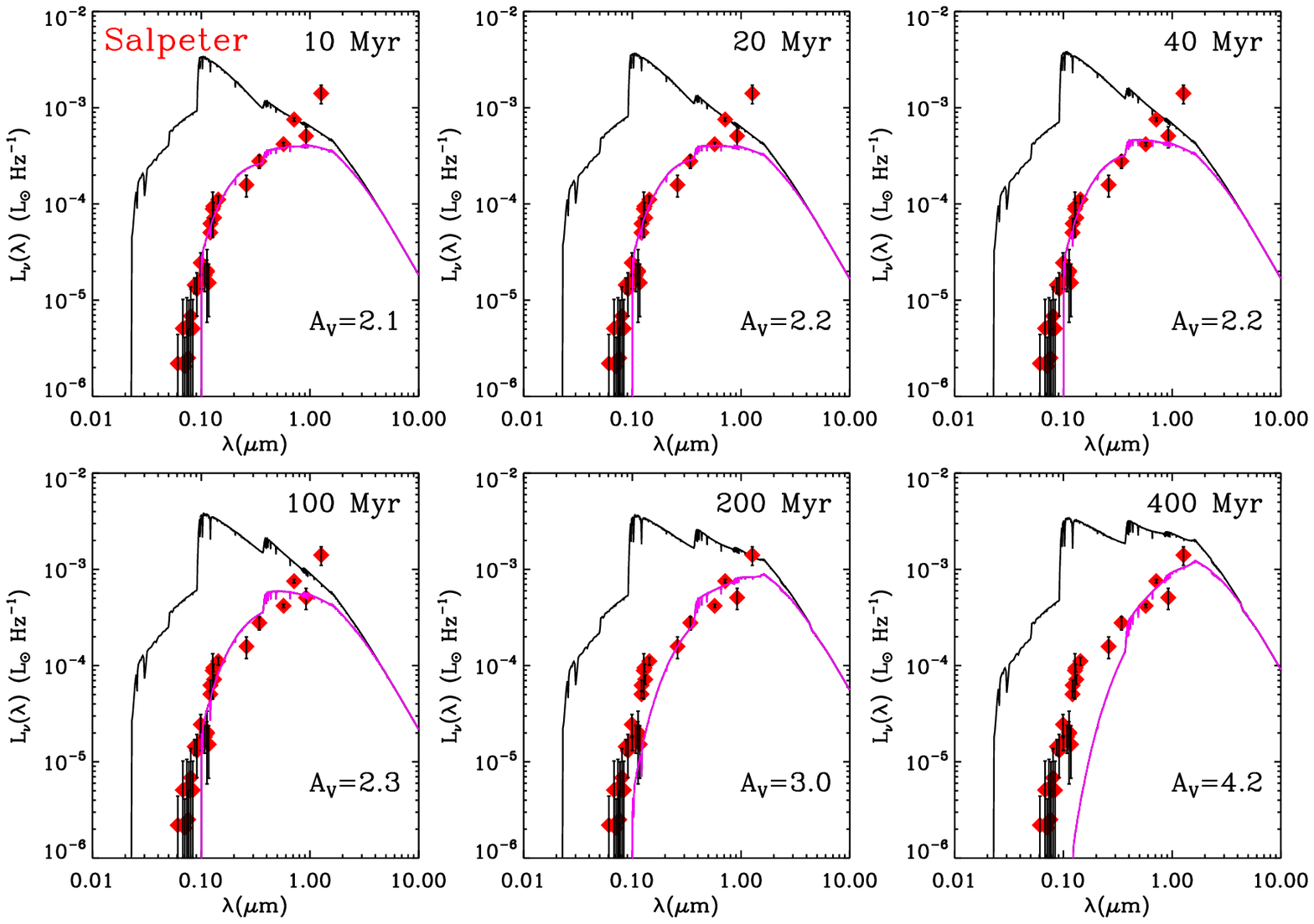}\\
    \vspace{0.3in}
    \includegraphics[width=5.0in]{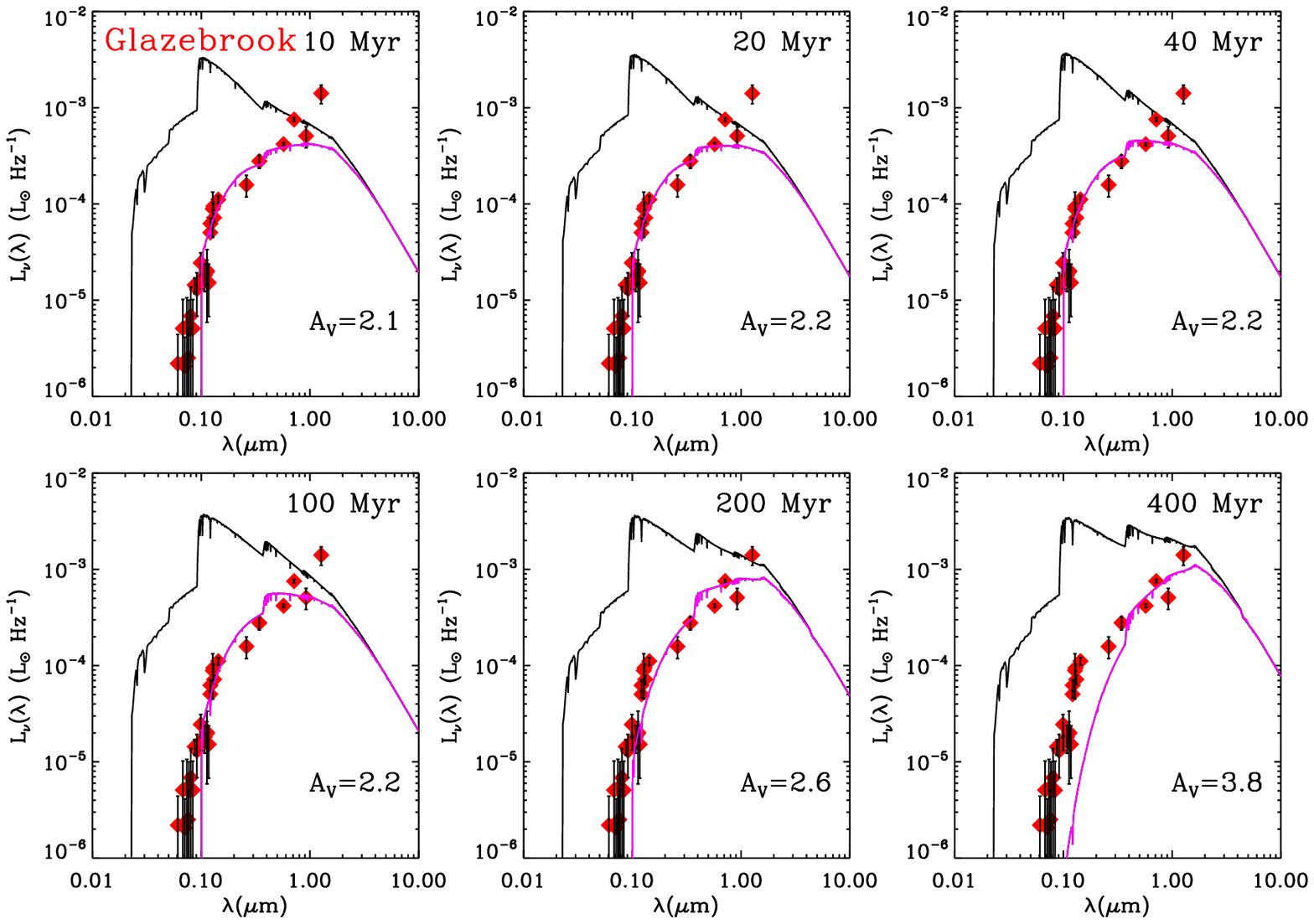}
  \caption{{\footnotesize
           Intrinsic (black curve) and attenuated (violet curve) stellar SEDs are plotted for an adopted $L_{bol} = 1\times10^{13}$~\lsun, and for 2 different IMFs (Salpeter and Glazebrook) and 6 different galactic ages. Each panel represents a distinct evolutionary scenario, characterized by a constant SFR determined by the requirement that the intrinsic bolometric luminosity be equal to the adopted observed value of $1\times10^{13}$~\lsun\ at the designated galactic age (see Figure~\ref{sfr}). The value of $A_V$ that provided the best fit of the attenuated spectrum to the observed UVNIR fluxes (red diamonds) is given in the figure. }}
\label{sed1}
\end{figure*}

\begin{figure*}[p]
  \centering
  \includegraphics[width=5.0in]{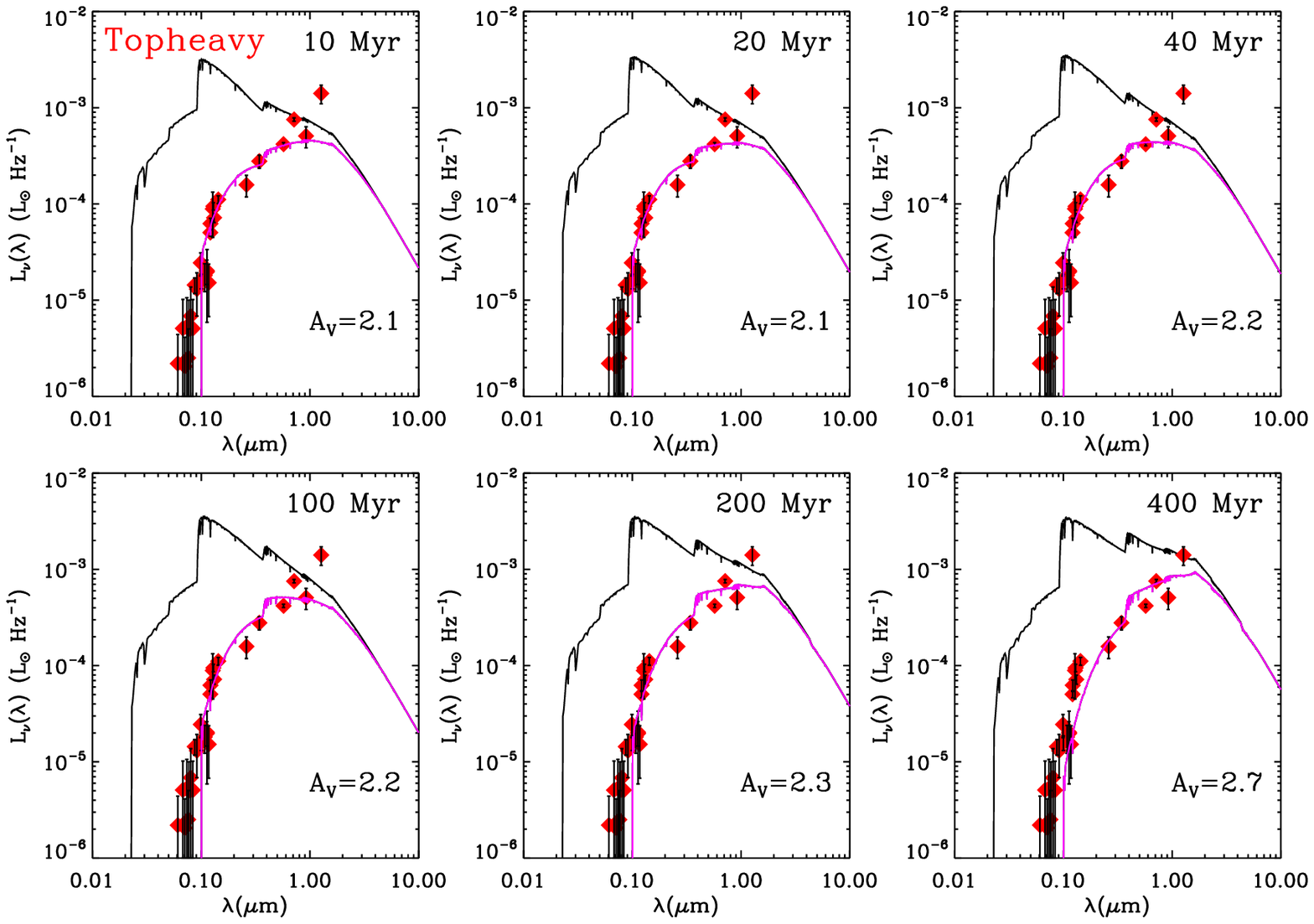}\\
  \vspace{0.3in}
    \includegraphics[width=5.0in]{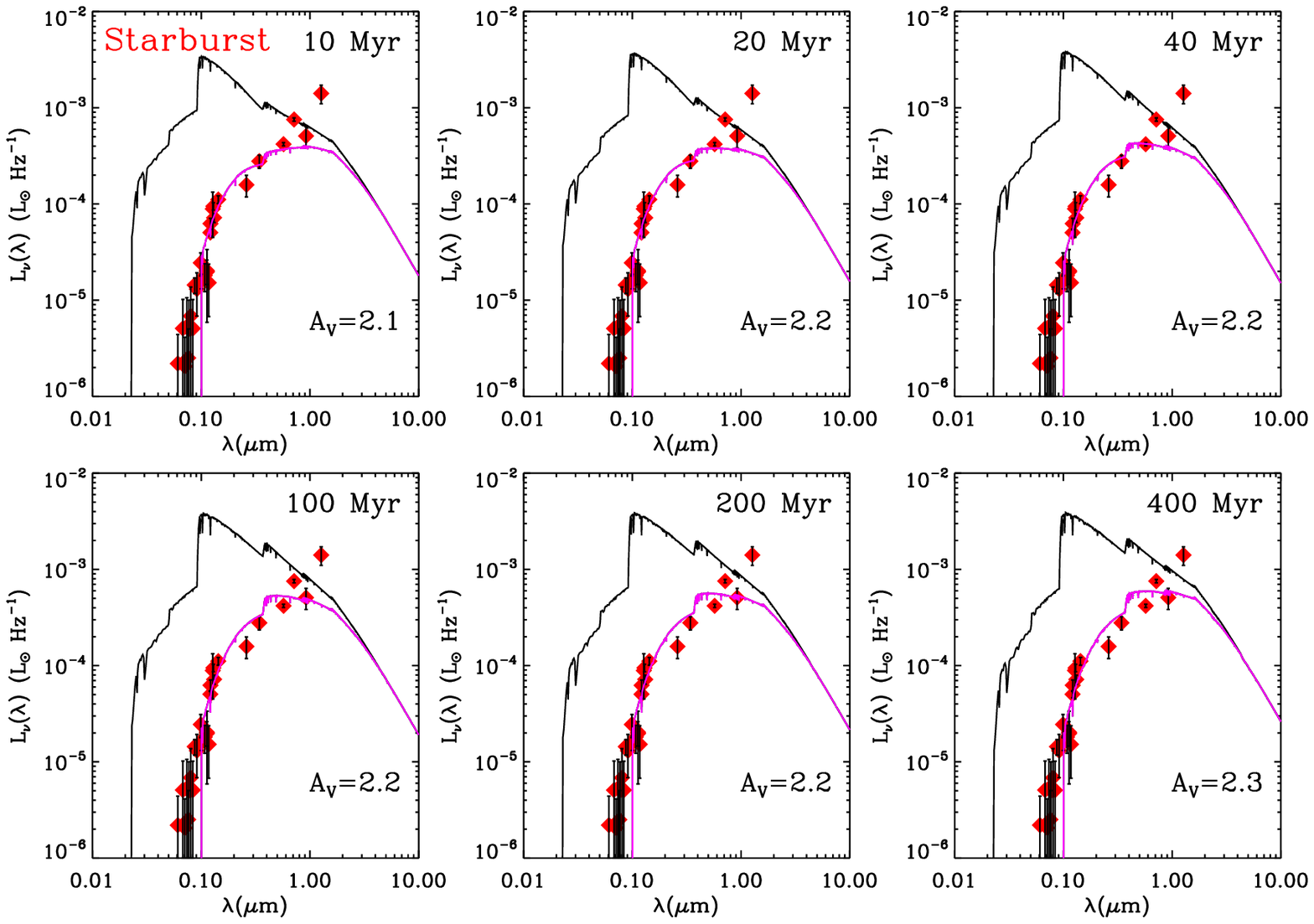}
  \caption{{\footnotesize
           Same as Figure~\ref{sed1} for the Top Heavy and Starburst IMFs.}}
\label{sed2}
\end{figure*}

The UVNIR SED of a galaxy can provide independent constraints on its age. Using the results of the previous section we created a library of intrinsic stellar SEDs for the seven IMFs and for a grid of galaxy ages, ranging up to to 800~Myr. The results are shown in Figures~\ref{sed1} and \ref{sed2} for 4 select IMFs and for ages between 10 and 400~Myr. 
Each panel in the figure presents the intrinsic SED (black line) generated by a distinct evolutionary sequence that is terminated at the assumed galactic age presented in the panel. Each evolutionary sequence is characterized by a constant SFR which is distinct from other sequences and determined by the stellar IMF and the galactic age for the adopted bolometric luminosity of  $1\times10^{13}$~\lsun\ (see Table~\ref{lumSB_sfr} and Figure~\ref{sfr}). So the Salpeter 20~Myr panel represents the SED of a galaxy that underwent a constant SFR of $\sim 1900$~\myr\ for that length of time. In each panel, the intrinsic SED is clearly in excess of the observed stellar flux (red diamonds), which has been significantly attenuated by dust. Lacking a definitive dust model for the IR emission, we used the Calzetti law \citep{calzetti00} to characterize the extinction. For each evolutionary sequence we calculated the visual optical depth, $\tau(V)$, that provided the best least-squares fit to the observations. Values of $\tau(V)$ ranged from $\sim 2-4$. The attenuated stellar spectra are presented by violet lines in the figures.

\begin{figure}[t]
  \centering
  \includegraphics[width=3.5in]{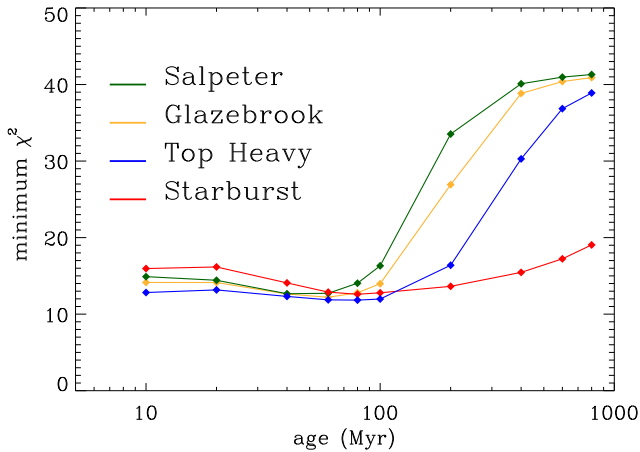}
  \vspace{0.1in}
  \caption{{\footnotesize
           The value of $\chi^2$ for the fit of the attenuated SEDs to the UVNIR observations for 4 different IMFs as a function of galactic age. The figure shows that for each IMF the best fitting SED is not sensitive to age. For ages less than $\sim 100$~Myr, the small changes in the intrinsic SED are compensated for by small changes in the amount of dust attenuation. The galaxy's SED provides only a weak limit on the age. The starburst IMF produces equally good fits to the observed SED for all galactic ages.}}
\label{chi2min}
\end{figure}

Figure~\ref{chi2min} depicts the minimum $\chi^2$ of the fit of the attenuated spectra to the data as a function of galaxy age and stellar IMF. The figure shows that for each IMF there exists an age degeneracy for which the value of $\chi^2$ is essentially unchanged. This degeneracy is caused by the fact that the galaxy's SED changes very little with time for the first $\sim 100$~Myr. The small changes that do occur are compensated for with small changes in the magnitude of the attenuation. This effect is exhibited in both the figures of the SEDs and that depicting the value $\chi^2$ with galactic age. For example, Figure~\ref{chi2min} shows that the Salpeter and Glazebrook IMFs provide  equally good fits to the observed UVNIR SED for ages up to $\sim 100$~Myr, and the Top Heavy IMF for ages up to $\sim 200$~Myr. The fits become significantly worse thereafter, an effect that is also clearly seen in Figures~\ref{sed1} and \ref{sed2}. The Starburst IMF provides an equally good fit to the observed spectrum at all epochs. 
Better fits to the observed spectra at later epochs can probably be obtained if we adopted a different extinction law, or a time-dependent extinction law that reflects the changing dust composition with time. Such changes would need to be consistent with the far-IR emission as well. Such detailed investigation into the extinction law (i.e. optical properties of the dust, and the physical distribution of the dust with respect to the radiation sources) is currently beyond the scope of this paper.

 \subsection{Stellar Mass Constraints of the Age: A Partial Lifting of the Degeneracy.}
The ambiguity in stellar age can be partially lifted by considering the upper limits on the stellar mass in the galaxy. \cite{capak11} derived a stellar mass of $\sim 10^{10}$~\msun, using the Maraston library of stellar SEDs \citep{maraston05} and a galactic age of $\sim 30$~Myr. This age was inferred from the Keck~II spectrum, assuming that all the star formation occurred in a single burst. Since we have shown that the UV spectrum cannot constrain the age of \az, we prefer to use a weaker, but observationally determined, constraint on the stellar mass. Using kinematic CO data and limits on the angular size of the galaxy, \cite{riechers10} derived the values of \mgas, the mass of the gas (dominated by the molecular component), and \mdyn, the dynamical mass of the galaxy. The dynamical mass is given by \mdyn $\simeq$ \mgas + $M_{stars} + M_{DM}$, where $M_{stars}$ and $M_{DM}$ are, respectively, the mass in stars and dark matter. The  dynamical mass offers therefore a strict upper limit on the stellar mass. Adopting a gas fraction ($\equiv$ \mgas/\mdyn) of 0.5 \citep{riechers10}, we adopt an upper limit of $M_{stars} \simeq M_{gas} \simeq 5\times 10^{10}$\msun\ on the stellar mass of the galaxy.

 \begin{figure}[t]
  \centering
  \includegraphics[width=3.5in]{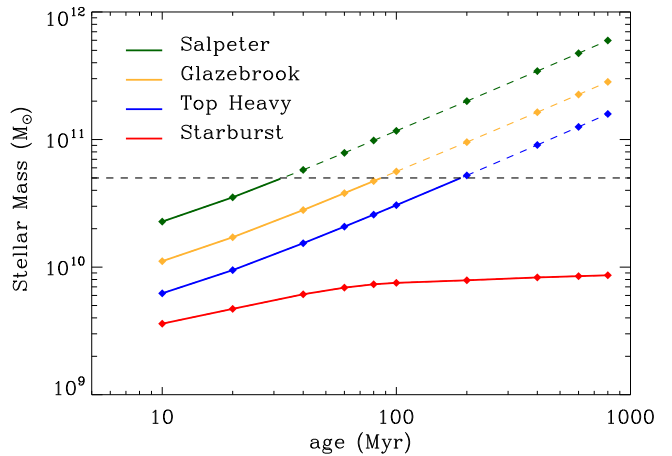}
  \vspace{0.1in}
  \caption{{\footnotesize
         The stellar mass as a function of the assumed duration of the star formation activity in AzTEC3 for four different IMFs. The dashed horizontal line corresponds to the upper limit on the stellar mass in the galaxy derived from estimates of the dynamical and gas masses of the galaxy \citep{riechers10}. For each IMF, the bold colored line depicts the allowed range of galaxy ages before the model violates the constraint on the stellar mass.  }}
\label{mstar}
\end{figure}

Since the mass of stars builds up at different rates for the different IMFs, this upper limit can provide useful constraints on the age of the galaxy.
This effect is shown in Figure~\ref{mstar}, which depicts the derived mass of stars as a function of the assumed age of \az\ for different stellar IMFs. Each symbol represents a distinct evolutionary sequence in which the SFR proceeded at a constant rate determined by the assumed galactic age and stellar IMF (see Figure~\ref{sfr}).    
The Salpeter IMF requires the largest SFR, and is therefore the first to exceed this mass limit after $t_{age}(max) \sim 30$~Myr. The Glazebrook IMF exceeds this mass limit after $\sim 80$~Myr, and the Top Heavy IMF after $\sim 200$~Myr. The Starburst IMF is the most efficient one in producing the observed bolometric luminosity with a low SFR, and does not have a large population of low mass stars. Consequently, it does not exceed the adopted stellar mass limit within $\sim 1$~Gyr, the approximate age of the universe at that redshift.

 \section{DUST EVOLUTION: A NEW CONSTRAINT \\ ON GALACTIC PROPERTIES}
The presence of $(2\pm 1)\times 10^9$~\msun\ of dust provides new additional constraints on the star formation rate and history of \az. To demonstrate this effect we used the chemical evolution model described in \cite{dwek11} to follow the evolution of the dust for the four different astrophysical scenarios determined by the stellar IMFs and the upper limit of the mass of stars in the galaxy. For each IMF we adopted a galactic age, \tmax, taken to be equal to the last epoch before the mass of stars produced by the model violates the stellar mass limit. The ages are $\sim 30$, 80, and 200~Myr for the Salpeter, Glazebrook, and Top Heavy IMFs, respectively (see Fig. \ref{mstar}). For the Starburst IMF we chose, quite arbitrary, a value of \tmax~=~800~Myr. The models then assume that star formation proceeds at a constant rate of $\sim 1800$, 700, 500, and 300~\myr, repectively  (see Fig. \ref{sfr}). 

 \begin{figure*}[t]
  \centering
  \includegraphics[width=3.0in]{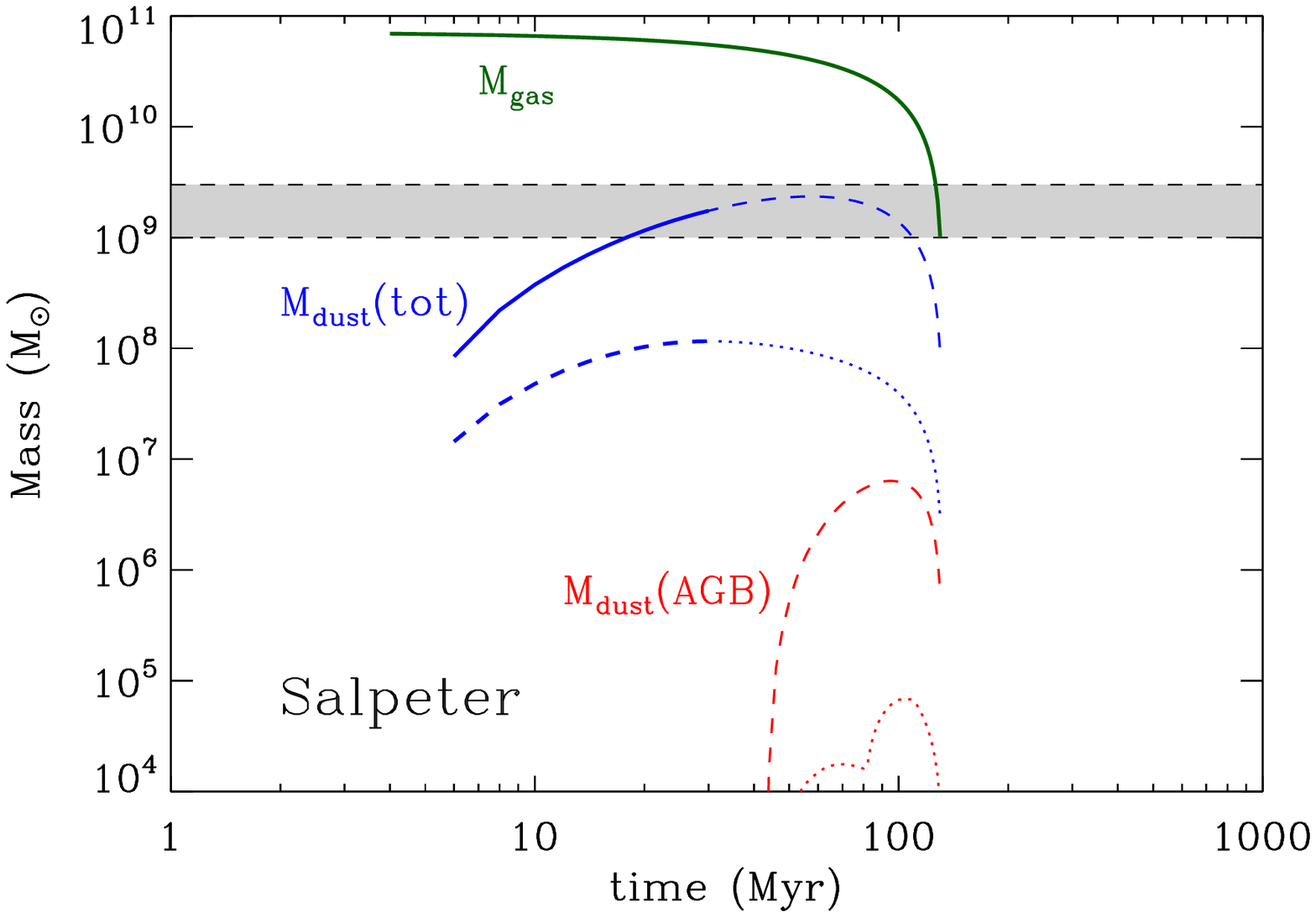} 
  \hspace{0.1in}
    \includegraphics[width=3.0in]{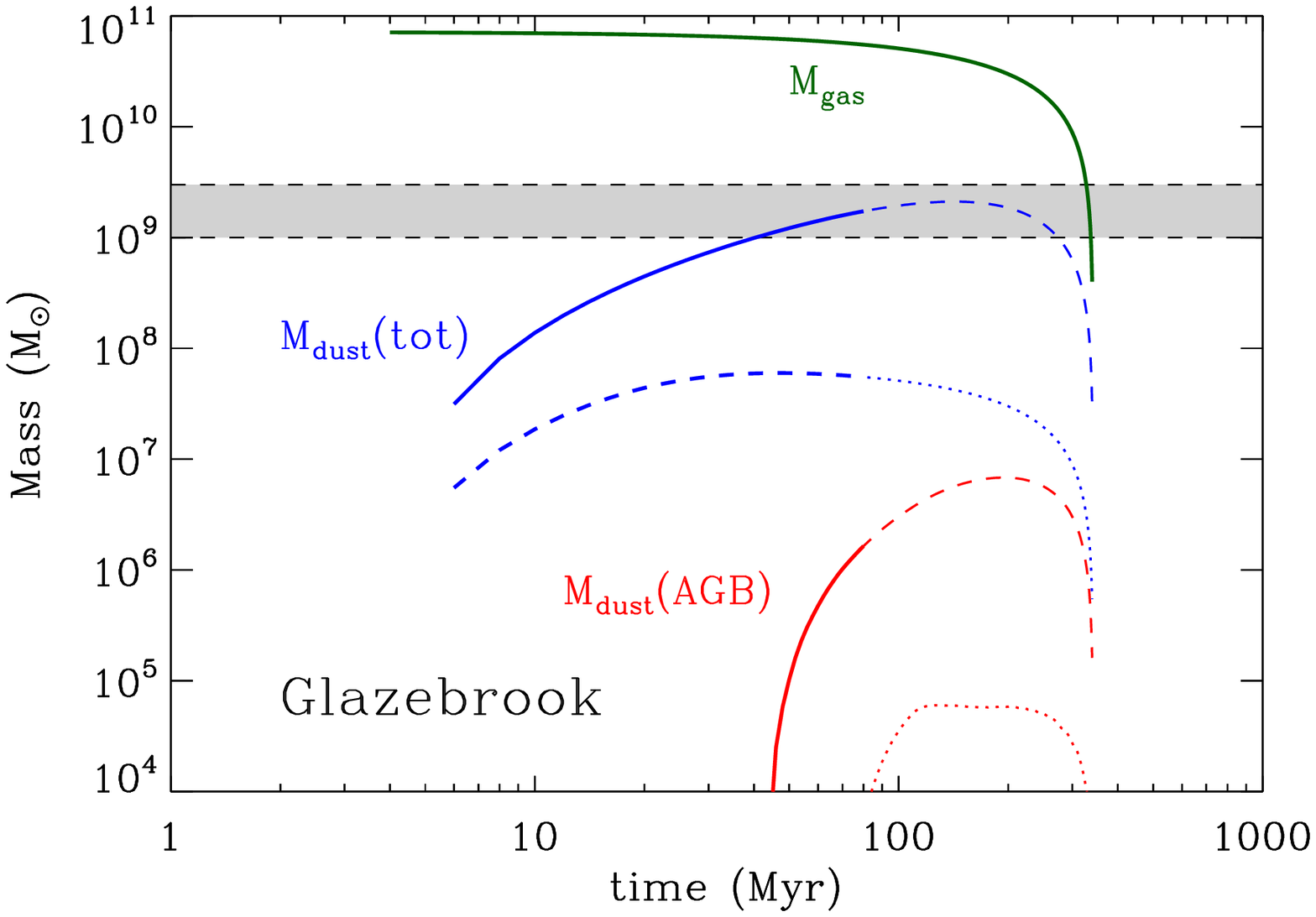}\\
   \vspace{0.2in}
      \includegraphics[width=3.0in]{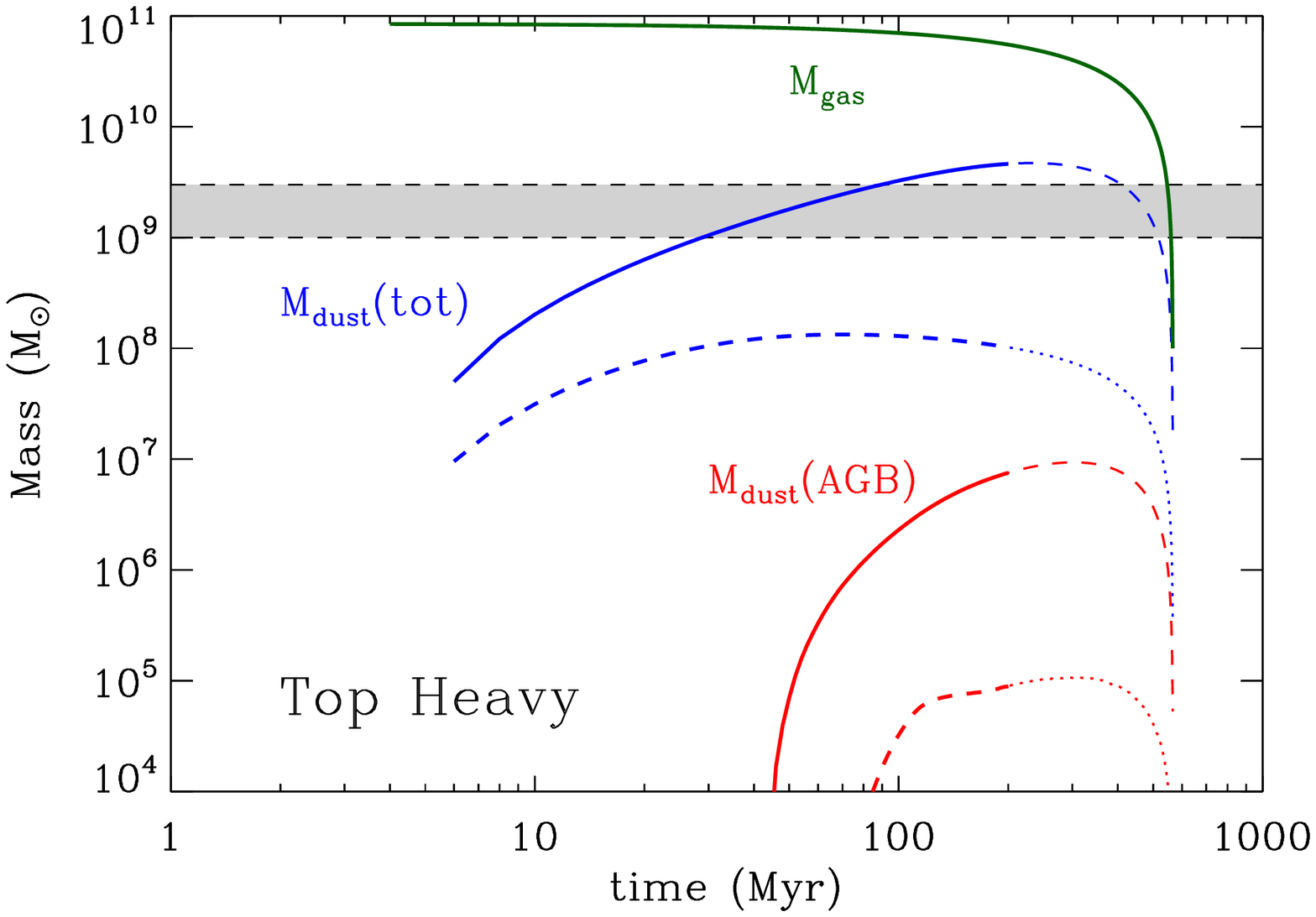} 
  \hspace{0.1in}
    \includegraphics[width=3.0in]{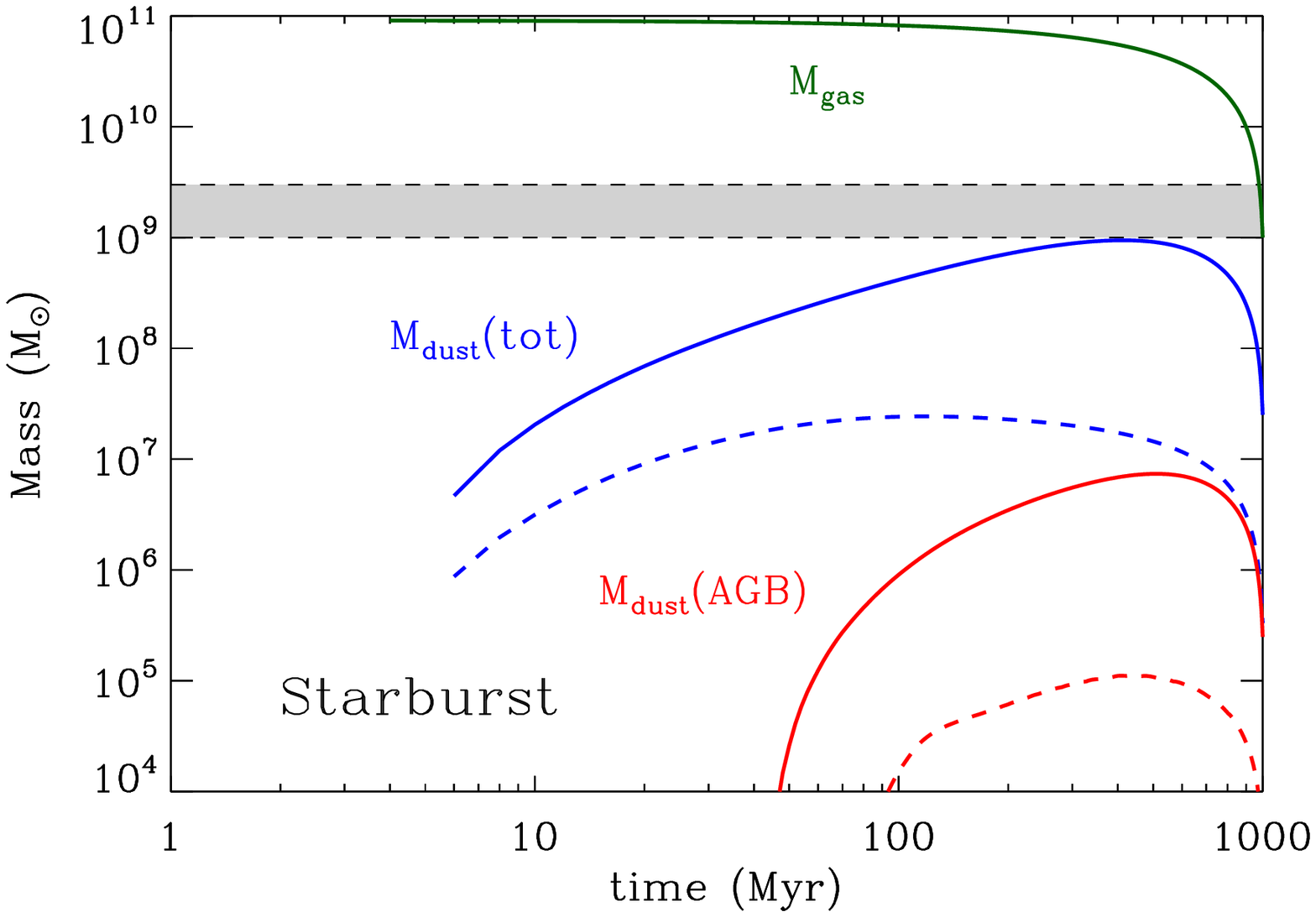}\\

  \vspace{0.1in}
  \caption{{\footnotesize The evolution of the dust and gas mass in \az\ are plotted as a function of time for the models characterized by their different IMFs. Total dust masses (blue curves) represent the sum of the dust masses produced by SNe and AGB stars. The contribution of AGB stars is depicted by red lines. Total and AGB dust masses are plotted for two cases: upper curves ignore the effects of grain destruction, whereas the lower curves adopt a Milky Way destruction efficiency. The bold section of each curve depicts the evolution over the time span allowed before the galaxy violates the upper limit on the stellar mass (see Fig.~\ref{mstar}). The shaded region depicts the adopted dust mass limits inferred from the far-IR observations.}}
\label{dustvol}
\end{figure*}  

We used a closed box model to follow the chemical evolution of \az. The initial gas mass was chosen so that each model reproduced the inferred gas mass of $\sim 5\times10^{10}$~\msun\ at age $t=$\tmax. The chemical evolution model follows the changes in metallicities over timescales that are comparable to the main sequence lifetimes of the massive SN progenitors. Consequently, in addition to the delayed recycling of matter by long-lived AGB stars, the model also includes the delayed recycling of matter by massive stars. AGB yields for stars with masses below 8~\msun\ were taken from \cite{karakas07} for a metallicity of $Z=0.008$. SN dust yields of massive stars were  taken as an average between the zero and solar metallicity yields given by \cite{heger10} and \cite{woosley07}, respectively (see Table~\ref{dust_yield} for select stellar masses). Dust condensation efficiencies were chosen to be unity for AGB stars, and 0.5 for Type~II SNe.

The results of our model calculations are shown in Figure~\ref{dustvol}. The green curve depicts the evolution of the gas mass which plummets when the constant SFR depletes all the gas (and dust) into stars. The blue curves depict the evolution of the total dust mass, which is the sum of the contribution from SNe and AGB stars. The AGB contribution is a small fraction of the total dust mass, and is shown separately by red curves. Models were run for two cases: the first ignores the effect of grain destruction, and is depicted by the solid blue curves; and the second takes the effect of grain destruction into account and is depicted by the dashed blue curves. The grain destruction rate was calculated by integrating the mass of dust destroyed as a function of shock velocity \citep{jones04}, over the evolution of a supernova remnant. The mass of dust destroyed by a single SN remnant during its lifetime is about $m_{dest} \approx 3$~\msun\ \citep{dwek07b} and the grain destruction rate is given by $SFR\times m_{dest}/m_{\star}$.  The bold section of each curve depicts the evolution over the time span allowed before the galaxy violates the upper limit on the stellar mass (see Fig.~\ref{mstar}). The shaded region depicts the range of dust masses inferred from the far-IR observations.

The figure shows that when grain destruction is taken into account, none of the models is capable of producing the required dust mass at any time. 
With no grain destruction, all models depicted in the figure are capable of producing the required dust mass, albeit at  different quantities. These idealized models can be considered as representatives of models in which rapid accretion in molecular clouds compensates for the mass of dust that is destroyed in the ISM. 
So the models show that rapid grain growth in clouds is an essential process for producing the inferred dust mass in \az.

The model that is clearly most capable of producing the inferred dust mass is the one with the Top Heavy IMF. At the age of 200 Myr it produces more than the inferred upper range of the dust mass, thus allowing for a modest amount of grain destruction. It therefore does not require that the accretion in molecular clouds reconstitute all the dust that was destroyed in the more diffuse ISM. The Starburst model is just capable of producing the lower range of inferred dust mass at $t=400$~Myr, and requires that {\it all} the refractory elements returned by grain processing into the gas  are quickly reaccreted onto surviving grains in molecular clouds. The models with a Salpeter or Glazebrook IMF fall in between these two models.

 \begin{figure}[t]
  \centering
  \includegraphics[width=3.5in]{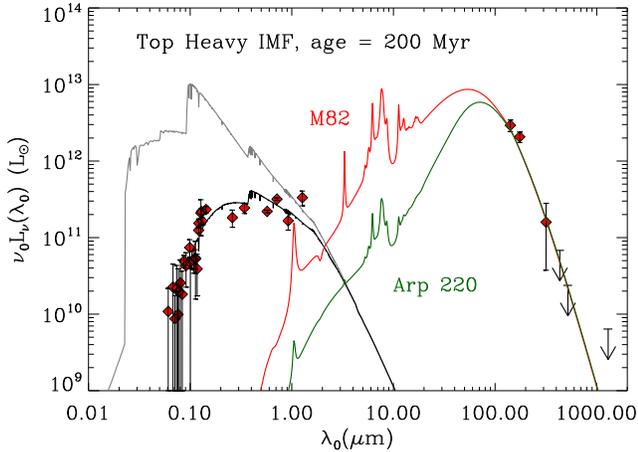} 
   \caption{{\footnotesize The intrinsic UV to far-IR luminosity from \az\ (grey line) calculated with the P\'EGASE stellar population code is plotted as a function of the rest frame wavelength, $\lambda_0$. The two galactic templates of M82 and Arp220 represent two possible fits to the far-IR spectrum of AzTEC3. The intrinsic stellar luminosity of $10^{13}$~\lsun\ falls between that of the two galactic templates (see Table~\ref{dust_tml}). The black curve represent the attenuated starlight derived with the Calzetti extinction curve for a value of $A(V)=2.30$. }}
\label{spec2}
\end{figure}  

Figure~\ref{spec2} depicts the spectrum produced by stars generated by the most plausible scenario that satisfies both the stellar mass and the dust mass constraints of \az: a galaxy with a Top Heavy IMF undergoing a constant SFR at a rate of 300~\myr\ over a timespan of 200~Myr, and allowing for over half of its dust to be destroyed. The stellar SED is attenuated by a Calzetti extinction law with $\tau(V)=2.3$. The intrinsic stellar spectrum is shown as the gray curve, and has a luminosity of $1\times 10^{13}$~\lsun, which lies between the luminosity of the two galactic templates, M82 and Arp~220 (see Table~\ref{dust_tml}). This scenario is not unique, but the most plausible one among the models. A Top Heavy IMF was also the preferred scenario for the rapid and efficient production of dust in the early universe by \cite{gall11a} and \cite{dunne10}.

 \section{A POPULATION III ORIGIN FOR THE DUST?}

\begin{deluxetable*}{lcccccc}
\tablewidth{0pt}
\tablecaption{Yields of Condensible Elements and Maximum Dust Masses in Massive Stars\tablenotemark{1}}
\tablehead{
\colhead{Element/} &
\colhead{ } &
\multicolumn{2}{c}{20 \msun} & 
\colhead{ } &
\multicolumn{2}{c}{50 \msun} \\
\cline{3-4}\cline{6-7}
\colhead{Metallicity} &
\colhead{ } &
\colhead{$Z=0$} &
\colhead{$Z=Z_{\odot}$} &
\colhead{ } &
\colhead{$Z=0$} &
\colhead{$Z=Z_{\odot}$} 
 }
\startdata
$^{12}$C		&	& 0.21 		&  0.23	&	& 1.86 &  1.61 \\
$^{16}$O		&	& 1.34 		&  1.38	&	& 11.2 &  9.7 \\
$^{24}$Mg	&	& 0.065 		&  0.065	&	& 0.31 &  0.33 \\
$^{28}$Si	&	& 0.0025 	&  0.19	&	& 0.27 &  0.21 \\
$^{56}$Fe	&	& $\sim 0$ 	&  0.12	&	& 0.41 &  0.28 \\
\hline
Carbon dust 						&	& 0.21 			&  0.23		&		& 1.86 		& 1.61 		\\
Silicate dust\tablenotemark{2}	&	& 0.11 			&  0.52		&		& 1.10 		& 1.00 		\\
Iron dust 						&	& 0.0 			&  0.12		& 		& 0.41 		& 0.28 		\\
{\bf Total dust}\tablenotemark{3}				&	& {\bf 0.32} 	& {\bf 0.87}	& 		& {\bf 3.4}  & {\bf 2.9} \\
\enddata
\tablenotetext{1}{Zero metallicity yields were calculated for models with no mixing and with explosion energies of $1.2\times 10^{51}$~erg and $5\times 10^{51}$~erg, for the 20 and 50~\msun\ stars, respectively \citep{heger10}; \zsun\ metallicity yields were taken from \cite{woosley07}. All entries are in units of \msun.}
\tablenotetext{2}{The silicate dust mass was calculated assuming that all the Mg and Si are locked up in MgO and SiO$_2$ dust precursors.}
\tablenotetext{3}{Total dust masses listed assume a 100 percent condensation efficiency.}
\label{dust_yield}
\end{deluxetable*}
Population~III (Pop~III) stars, the first population of stars being born out of a pristine, zero metallicity gas, may also be important contributors to the dust in these young galaxies \citep{nozawa03, nozawa09,todini01, cherchneff10}. They are believed to be massive \citep{bromm04}, and supported by radiation pressure at the Eddington luminosity: 
\begin{equation}
\label{1}
\left({L_{edd}\over L_{\odot}}\right)=3\times10^4\, \left({m \over M_{\odot}}\right).
\end{equation}

Consider a population of massive Pop~III stars all of identical mass $m$, and let \lbol\ be the total bolometric output of this stellar population. To maintain a steady abundance of such stars, they have to be created at a rate given by:
\begin{equation}
\label{dndt}
{dN\over dt} = \left[{L_{bol} \over L_{edd}}\right]\times \tau_{MS}^{-1} \qquad ,
\end{equation}
where $\tau_{MS}$, their main sequence lifetime, is determined by the time it takes to radiate all the energy released in the production of helium and heavy elements at the Eddington luminosity. It is independent of stellar mass, and given by: 
\begin{equation}
\label{2}
\tau_{MS}=0.007 mc^2/L_{edd} \approx 3\times10^6 \ {\rm yr}.
\end{equation} 
The SFR needed to maintain the observed luminosity is independent of the stellar mass and given by $SFR=m\, (dN/dt) \approx 10^{-12}\times L_{bol}$~\myr.
The rate of dust production is given by:
\begin{equation}
\label{ pop3}
{dM_{dust}\over dt} = Y_{dust}\, \left({dN\over dt}\right) = \left({Y_{dust}\over \tau_{MS}}\right)\ \left({L_{bol} \over L_{edd}}\right) \qquad
\end{equation}
where $Y_{dust}$ is the amount of dust produced in these objects.

Dust formation in massive Pop~III stars has only been calculated for 170~\msun\ stars \citep{nozawa03, cherchneff10}. Such a Pop~III star will have a luminosity of $\sim 5\times 10^6$~\lsun. The dust yield depends on the whether the ejecta maintains its stratified compositional structure or becomes completely mixed, and ranges from $\sim 7$ to $\sim 35$~\msun, for unmixed and fully mixed ejecta, respectively.  
The total mass of dust produced at time $t$, assuming no grain destruction, is given by:
\begin{equation}
\label{mdust}
\left({M_{dust}\over M_{\odot}}\right)\approx (5- 25)\times \left({t\over yr}\right)\qquad 
\end{equation} 
where the range of ages reflects the uncertainty in the dust yield of such stars.
The time needed to create an average $M_{dust}=2\times10^9$~\msun\ of dust by 170~\msun\ Pop~III stars is therefore $\sim 100-400$~Myr. This timescale is significantly longer than their main sequence lifetime, and too long to maintain a pristine reservoir of gas to feed the formation of metal-free stars. 
We therefore conclude that the massive amount of dust seen in \az\ cannot have been produced by a burst of Pop~III stars that are currently giving rise to the observed far-IR luminosity of the galaxy.

 \section{DISCUSSION}
Deriving the physical properties of high redshift galaxies from their spectra is an important goal for understanding their origin and evolution. 
In the absence of perfect and complete information, these derived physical properties are not unique, allowing for different evolutionary scenarios to explain the observations. 

In this paper we explore a new method that can, in principle, discriminate between  different evolutionary scenarios. The method requires that the mass of dust inferred from the far-IR emission to be present in these galaxies be produced within an allotted time span imposed by other observations. 
We applied our method to the analysis of the ultraluminous submillimeter galaxy \az. The basic observations available to constrain the models are the high-resolution UV-optical spectrum, the UV to millimeter SED, and the CO transitions that were detected from this object. 

A fundamental observed quantity is the bolometric luminosity of the galaxy which in \az\ is dominated by the far-IR emission, and can be used to infer the SFR of the galaxy. However, converting \lbol\ to a SFR requires knowledge of the stellar IMF and the age of the galaxy, since for a constant SFR and given IMF, the bolometric luminosity increases with time because of the contribution of long-lived low mass stars to the emission. 

The IR region of the spectrum can be used to derive the mass of dust giving rise to the emission. The dust mass depends on the dust composition and the dust temperature distribution, both unknown quantities. Estimates of the dust mass can be derived for pure dust compositions (graphite, amorphous carbon, or silicates), based on dust models for the Milky Way. Alternatively, the dust mass can also be estimated by fitting the observed SED with that of known galaxies for which the dust mass has been derived from detailed models. 

The CO emission spectrum  can be used to derive the dynamical mass of the galaxy and the mass of molecular gas.
Both quantities are uncertain. The dynamical mass estimate depends on the basic premise that the system has relaxed to a Keplerian disk, and the determination of the gas mass depends on the uncertain X-factor that converts the CO line intensity to an H$_2$ mass.
Without any further assumptions, only an upper limit can be obtained for the stellar mass in this galaxy.

The high-resolution UV spectrum can be used to determine the age of the system. Massive young stars exhibit characteristic P-Cygni profiles. The presence of these features in the spectrum can therefore be used to determine the age of the system. For an instantaneous burst of star formation, these features disappear as these massive stars die off within $\sim 30$~Myr. So their presence can be used to estimate the age.  When star formation is a continuous process, the massive stars are always present, and the strengths of the features relative to the continuum changes then much more gradually. Therefore, the age inferred from such spectrum will be much longer than that inferred for an instantaneous burst. From our simulations, assuming a constant SFR, we find that the UV spectrum cannot discriminate between systems that are between $\sim 30$ to 400~Myr old, the relevant age limits for \az.

The bolometric luminosity of \az\ lies between 0.6 to $2\times10^{13}$~\lsun, and to be definitive we adopted a value of \lbol $= 1\times 10^{13}$~\lsun. For the dust mass we adopted a value of  $2\pm 1\times 10^9$~\msun.
We adopted a very conservative upper limit of $5\times10^{10}$~\msun\ on the stellar mass, based on the estimated dynamical and gas mass from this object. 

Using these basic constraints, \lbol, $M_{dust}$, and $M_{stars}$, we constructed  models to follow the evolution of the galaxy's SED, stellar, and dust content as a function of time for 7 different stellar IMFs.
Viable models were those that were able to produce the inferred mass of dust before the concurrently produced mass in stars exceeded the mass limit.
The results of these models are depicted in Figure~\ref{dustvol}, and described in Section 4.

The results of the paper can be briefly summarized as follows:
\begin{enumerate}
  \item Most of the dust giving rise to the far-IR emission from \az\ must have been grown in molecular clouds. 
  In the Milky Way, the need for grain growth in molecular clouds to explain the abundance of dust was first pointed out by \cite{dwek80b}, and more recently by \cite{zhukovska08} and \cite{jones11}. \cite{michalowski10a} argued that growth in molecular clouds is the main source of dust in high-redshift ($5 < z < 6.5$) quasars. In this paper we have extended this point to the ultraluminous starburst \az\ using detailed chemical evolution models.
  \item There is a correlation between grain destruction and the required growth in molecular clouds. At the peak of its dust production, which occurs at $\sim 200$~Myr, the Top Heavy model can form the inferred dust mass in \az\ without resorting to grain growth in the ISM, but with a significantly reduced (compared to the Milky Way) efficiency of grain destruction. At earlier epochs grain growth in molecular clouds must account for a significantly larger fraction of the dust mass. In the Starburst model any grain destruction must be rapidly balanced by accretion in the ISM. The Salpeter and Galzebrook IMFs are less efficient in producing dust than the Top Heavy one. They can afford only little grain destruction, and therefore require a large fraction of their dust to be of interstellar origin. 
  \item The model with the Top Heavy IMF is the most efficient one for producing the dust in \az, and therefore the more favorable model for  the origin of the dust in this galaxy. Taking the galaxy's age to be 200~Myr, the longest time allowed by the stellar mass constraint, the constant SFR required to provide the observed bolometric luminosity and the required dust mass is $\sim 500$~\myr. 
  The inferred constant SFR is significantly smaller and the age is significantly longer than the previous estimates of $\sim 1800$~\myr and $\sim 30$~Myr, respectively, assuming a single burst of star formation. 
    \item The Top Heavy model is, however, not unique. A clear discrimination between the different models requires a significant reduction in the uncertainties of the measured quantities. For example, a better determination of the stellar mass may have resulted in a lower value than the upper limit adopted in this paper, ruling out the models using the Salpeter or the Glazebrook functions to characterize the stellar IMF. 
  \item Finally, Population~III stars could not have produced both the currently observed luminosity and  dust mass in \az.
\end{enumerate}

The studies presented here illustrate the power of multiwavelength observations and the simultaneous use of spectral, stellar, and dust mass constraints for determining the properties of high-redshift galaxies. Our studies also highlight the need for more observational data and improved model input parameters.

{\bf Acknowledgements}
ED acknowledges helpful discussions with Dominik Riechers. GISMO millimeter observations used in the analysis were supported through NSF grants AST-0705185 and AST-1020981. We also acknowledge useful comments by the referee that helped clarify the paper.

~\\~\\~


\bibliographystyle{/Users/edwek/Library/texmf/tex/latex/misc/aastex52/aas.bst}
\bibliography{/Users/edwek/science/00-Bib_Desk/Astro_BIB.bib}


%

\end{document}